\documentstyle[12pt]{article}
\input{epsf.tex}

\begin{document}
\centerline{\Large\bf Unitary theories in the work of Mira
Fernandes }
\smallskip
\centerline{\Large\bf (beyond general relativity and 
differential geometry)}

\bigskip

\centerline{\large\bf Jos\'e P. S. Lemos}
\centerline{\small Centro Multidisciplinar de {}Astrof\'{\i}sica -
CENTRA,}
\centerline{\small Departamento de F\'{\i}sica, Instituto Superior 
T\'ecnico - IST,}
\centerline{\small Universidade T\'ecnica de Lisboa - UTL,} 
\centerline{\small Avenida Rovisco Pais 1, 1049-001 Lisboa, Portugal,}
\centerline{\small email: joselemos@ist.utl.pt \footnote
{Published in {\it Boletim da Sociedade Portuguesa de Matem\'atica},
{\it (N\'umero Especial - Aureliano Mira Fernandes)}, eds.~L. Saraiva e
J. T. Pinto, (Sociedade Portuguesa de Matem\'atica, Lisboa, 2010),
p. 147. Based on the invited talk at the conference ``Mira Fernandes
and his age - An historical Conference in honor of Aureliano de Mira
Fernandes (1884-1958)'', Instituto Superior T\'ecnico, Technical
University of Lisbon, June 2009.}
}
\vskip 0.8cm
\centerline{\small Abstract}
\vskip 0.2cm
\vbox{\noindent \small
An analysis of the work of Mira Fernandes on unitary theories is
presented. First it is briefly mentioned the Portuguese scientific
context of the 1920s. A short analysis of the extension of Riemann
geometries to new generalized geometries with new
affine connections, such as those of Weyl and Cartan, is given.  Based
on these new geometries, the unitary theories of the gravitational
and electromagnetic fields, proposed
by Weyl, Eddington, Einstein, and others are then explained.  Finally,
the book and one paper on connections and two papers on unitary
theories, all written by Mira Fernandes, are analyzed and put in
context.}

\vskip 3mm

\noindent

\vskip 1cm

\section{Introduction}
\label{Intro}

\subsection{\small\bf Mira Fernandes background and 
the Portuguese context}

Aureliano de Mira Fernandes, born in 1884 in Portugal, was professor
of differential and integral calculus, rational mechanics, and other
lecture courses in mathematics, at Instituto Superior T\'ecnico {\tiny
(IST)}, from its foundation in 1911, onwards, until his
retirement. IST was situated provisionally at Rua do Instituto
Industrial (by Rua Conde Bar\~ao) near the river Tejo. Provisionally
means until the end of the 1930s, when IST moved to the place where it
is now, in the middle of the town.  He was also professor of
mathematical analysis at the Instituto Superior de Ci\^encias
Econ\'omicas e Financeiras {\tiny (ISCEF)}, what is now the Instituto
Superior de Economia e Gest\~ao {\tiny (ISEG)}.

Mira Fernandes has a vast work in theoretical physics and mathematics, 
his complete works have now been 
published \cite{gulb1,gulb2,gulb3}.
Mira Fernandes, by formation, was a mathematician not a physicist.
His Doctoral dissertation in 1911, supervised by Sid\'onio Pais, 
on ``Galois theory'', was submitted when he was 27 years old
\cite{Miradissertation} (see also \cite{rica}).  From his
dissertation to 1924, when he was 40 years old, there are no
publications. From 1924 onwards there are many publications on
several subjects, namely, group theory, differential geometry, unitary
theories and rational mechanics. There is no direct explanation for
this 13-year gap in publications, the only reasonable one is that
during those 13 years he was busy in preparing the lectures he had to
deliver as well as acquainting himself with the new subjects he was
interested.  The most important papers were published in
Rendiconti della Accademia dei Lincei, due to his friendship
with Levi-Civita, the great Italian mathematical physicist. After
Levi-Civita's compulsory retirement, Mira Fernandes published mainly
in Portuguese journals.

He corresponded heavily with Levi-Civita (see \cite{tazzioli}) and
also corresponded with \'Elie Cartan. Cartan in his work ``Les espaces
de Finsler'' \cite{cartan} writes (my translation) ``It is after an
exchange of letters with M. Aurelio (sic) de Mira-Fernandes, that I
have perceived of the possibility of this simplification''. This means
he had relations of value to him and to his country. He also 
corresponded with Portuguese mathematicians \cite{ccosta}.

At the time, in mathematics in Portugal, there was the towering figure
of Gomes Teixeira in Porto, a worldwide recognized mathematician with
works in the theory of curves and surfaces. Also in Porto, there was
Leonardo Coimbra, a philosopher who occasionally wrote on physics. In
Lisbon, Mira Fernandes had no peer. He was a member of the Lisbon
Academy of Sciences from 1928 onwards. In 1932 he proposed Levi-Civita
and Einstein to be foreign members of the Academy, a proposal
immediately accepted by the President of the Academy, Egas Moniz, the
future Nobel prize in medicine. These proposals by him were apt, since
these two figures were pioneers in differential and Riemann
geometry, general relativity and unitary theories, areas for which
Mira Fernandes devoted a great part of his scientific life.  In these
matters there was some interest by some community in Portugal,
although mostly dilettante, the exceptions being Ant\'onio Santos
Lucas, a professor in the Faculdade de Ci\^encias of Lisbon, who
delivered lectures on general relativity there, and perhaps some other
instances, although it seems that Eddington's expedition to Pr\'incipe
in 1919 to observe the light shift due to the gravitational field of
the sun, was not greeted with enthusiasm and interest by the
Portuguese scientific community. For the details of the scientific
context in Portugal in Mira Fernandes' time, see the excellent
studies in \cite{costaleitegagean,fitas} (see also 
\cite{fitasconf}).

\subsection{\small\bf Aim and plan of work}

In this article we will study Mira Fernandes works related to unitary
theories, what are now called theories of unification.  These theories
tried to unify the gravitational and the electromagnetic fields, the
two know fields at the time, into a single field.  Since these
theories are related to the theory of connections in differential
geometry, a theme that was dear to Mira, we also review his works on
the theory of connections.

The plan of the work is as follows. Above we outlined the scientific
context of the epoch in which Mira Fernandes was immersed. In section
2 we will outline the scientific context of the unitary theories of
the time, a time that spans from about 1916, the year of the creation
of general relativity, to about 1934, the year of the last work of
Mira Fernandes on the subject.  Some generalities related to general
relativity and Riemannn geometry, the geometry on which the theory is
based will be laid out.  The extension of Riemannn geometry to Weyl
geometry, the first unification scheme in this context proposed by Weyl
himself, and its potential development performed by Eddington, will be
revised. The extension of Riemannn geometry to include torsion given
by Cartan, will also be mentioned.  We will display the spectrum of
unitary theories based on the different geometries and connections of
the time, and mention the various unsuccessful attempts made by
Einstein to find the true unitary theory.  We will also refer to a
field, the $C$-field, which makes a bridge between the contravariant
and the covariant vectors and tensors.  Then in section 3 we will
delve into Mira Fernandes' works.  First we will analyze his works on
connections, namely, the book and the 1931 paper in Rendiconti dei
Lincei, and we will comment on them. Finally we will study his two
works of 1932 and 1933, also published in the same journal, which
apply the theory of connections to some unitary theories of
gravitation and electromagnetism.  In these works Mira Fernandes finds
an application for the $C$-field as the physical field of
electromagnetism in the unitary theory of Straneo. This is unique.  In
section 4 we conclude, commenting on other interesting works of Mira
in this conjunction, and on the fate of the unitary theories.
Finally, the main sources to write this article on history of unitary
theories and Mira Fernandes will be discussed, and the motivations for
writing it plus the acknowledgments will be given. The text is thus
divided as,
\begin{description}
\item[1.] Introduction
\begin{description}
\item[1.1] Mira Fernandes background and the Portuguese context
\item[1.2] Aim and plan of work
\end{description}
\item[2.] The scientific context of unitary theories
\begin{description}
\item[2.1] Generalities and general relativity (1916)
\item[2.2] Weyl geometry and Weyl 
theory of gravitation and electromagnetism (1918) 
\item[2.3] Eddington theory (1921)
\item[2.4] Cartan's torsion, differential geometry, Einstein's attempts
and the spectrum of unitary theories, and the $C$-field
\end{description}
\item[3.] The works of Mira Fernandes on connections and on unitary
theories of gravitation and electromagnetism
\begin{description}
\item[3.1] The work on connections (i) The book 1926 (ii) Rendiconti
1931
\item[3.2] Application to unitary theories of gravitation and
electromagnetism (i) Rendiconti 1932 (ii) Rendiconti 1933
\end{description}
\item[4.] Conclusions
\begin{description}
\item[4.1] What else? 
\item[4.2] The fate of unitary theories
\end{description}
\begin{description}
\item{Sources}
\item{Acknowledgments}
\end{description}
\end{description} 

\section{The scientific context of unitary theories}
\label{scientificcontext}

\subsection{\small\bf Generalities and general relativity (1916)}

The idea of unification in physics is an old one.  One of the first
attempts to unify fields and particles in the same scheme was provided
by Mie in 1912 \cite{mie}. In \cite{grhistorybook} 
an English  translation of the original paper, as well as 
of other important papers with a complete set of comments
is given.  An important follow up of this idea came later
through the work of Born and Infeld in 1934 \cite{borninfeld}, who
implemented this type of unification by modifying the Maxwell
Lagrangian, providing a non-linear extension with particle solutions
of the Maxwell equations. Nordstr\"om in 1914 \cite{nordstrom} tried
a different type of unification, not of fields and particles that
generate the fields themselves, but a unification of the different
fields. At the time there were two known fields, the gravitational and
the electromagnetic. In his attempt to unify his theory of
gravitation, a scalar one, with the Maxwell electromagnetic theory,
Nordstr\"om used a fifth spatial dimension, being thus the precursor
of the Kaluza-Klein theories.

The appearance of general relativity in 1916 \cite{gr}, inspired new
forms of unification. For instance Hilbert \cite{hilb} tried to use
Mie's ideas \cite{mie} in conjuction with general relativity to
produce a theory of particles and fields.  Soon, from its beautiful
structure based on Riemann geometry, general relativity would further
lead to many other unification schemes. To start with, general
relativity put the gravitational field in a special relativity
framework. However, it left electromagnetism out. Defining
$G_{\alpha\beta}$, $F_{\alpha\beta}$, $\tau_{\alpha\beta}^{\rm em}$,
and $j_\alpha$, as the Einstein tensor, the Maxwell tensor, the
electromagnetic energy-momentum tensor, and the electric current,
respectively, one may still argue that since the Einstein-Maxwell
equations lead to $G_{\alpha\beta}=8\pi \tau_{\alpha\beta}^{\rm em}$
and ${{F_\alpha}^{\beta}}_{;\,\beta}=j_a$ (Newton's constant $G=1$,
and the velocity of light $c=1$), there is a sense of
unification. This was put forward by Rainich in 1924 \cite{rainich}
and continued by Misner and Wheeler in 1954 \cite{misnerwheeler} in
what they called an already unified theory.  But those in pursue of
unification wanted more.

The argument for the unification went as follows, see Figure 1. 
The electric and magnetic fields had been unified into 
the electromagnetic field,
later shown that the whole unified scheme was consistent only using
special relativity and the corresponding spacetime arena.  Thus, one
might have argued, gravity (and so general relativity) and
electromagnetism, the two known fields of the time, should be
unifiable in a unitary theory using some special world background as
the correct arena. This was advocated by many, in particular by
Eddington, see \cite{eddingtonbook}.  What this special world
background could be was left imprecise. This rationale works if one
considers general relativity as a field theory, on the same footing of
electromagnetic theory. But even this is controversial. Is general
relativity a field or is it an arena as special relativity?

\vskip 1.0cm
\centerline{\epsffile{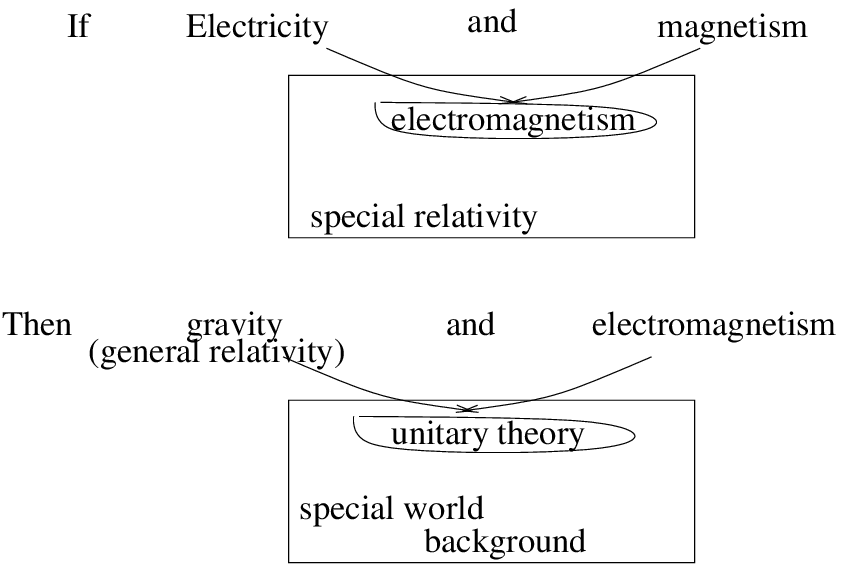}}
\noindent {\small
Figure 1. In this figure it is sketched the rationale that 
might have induced one to search for a unitary theory 
with its underlying special world background, in much the same 
way to what happened with electromagnetism and special relativity.}
\vskip 1.0cm

Here, we simply note that from 1916 onwards unification schemes have
always been forefront problems.

\subsection{\small\bf Weyl geometry and 
Weyl theory of gravitation and electromagnetism
(1918)}

The first attempt to unify gravitation and electromagnetism was
proposed by Weyl in 1918 \cite{weyl1918,weylbook}. In this theory the
electromagnetic potential is introduced as a geometrical quantity
which determines the transport law of a length scale.  The idea can be
decomposed into two parts. First, one has to develop a new geometry,
which in turn embodies the Riemann geometry, second one has to set
up a physical theory of gravitation and electromagnetism which in a
particular instance yields the Einstein-Maxwell equations.  Let us
analyze first the Weyl geometry, then the Weyl theory.

In Weyl geometry the transport of a vector $\xi$ with components 
$\xi^\alpha$, is given, as in Riemann geometry, by the equation
\begin{equation}
\delta \xi^\alpha=\Gamma^\alpha_{\beta\gamma}\,dx^\beta\,\xi^\gamma\,,
\label{transportlaw1}
\end{equation}
where $\delta \xi^\alpha$ is the change of the vector under transport, 
$\Gamma^\alpha_{\beta\gamma}$ is the connection, and $dx^\beta$
is a local displacement.  The difference to the Riemann geometry is
that the connection $\Gamma^\alpha_{\beta\gamma}$ is not given by the
Christoffel symbols $\{^{\,\,\alpha\,}_{\beta\gamma}\}$, 
composed of the metric alone, but is more
general.  The idea of a general transport, independent of the metric,
had been developed at about this time by Levi-Civita and others
\cite{levicivita1917} (see also \cite{scholz} for the ideas of Weyl in
relation to differential geometry and transport laws).  Now, comes the
new geometrical requirement.  At a point the length of a vector
$\xi^\alpha$ is given by $l\,^2=g_{\alpha\beta}\,\xi^\alpha\xi^\beta$,
where $g_{\alpha\beta}$ is a symmetric metric. For Weyl the length can
change under transport as
\begin{equation}
\delta l=\phi_\beta \,dx^\beta\,l
\label{transportlawlength}
\end{equation}
in analogy with equation (\ref{transportlaw1}), and where $\phi_\beta$
is a new field.  With the two requirements given by
Eqs.~(\ref{transportlaw1})-(\ref{transportlawlength}) one can deduce
after some algebra, (see, e.g., \cite{adler}) two things. One, that
the Weyl connection $\Gamma^\alpha_{\beta\gamma}$ should be given in
terms of the metric $g_{\alpha\beta}$ and the field $\phi_\beta$ as
\begin{equation}
\Gamma^\alpha_{\beta\gamma}=\{^{\,\,\alpha\,}_{\beta\gamma}\}
+g^{\sigma\alpha}\left(g_{\sigma\beta}\,\phi_\gamma+
g_{\sigma\gamma}\,\phi_\beta-g_{\beta\gamma}\,\phi_\sigma\right)\,.
\label{Gammas}
\end{equation}
The other is that the covariant derivative of the metric 
$g_{\mu\nu\,;\alpha}$ is given by 
\begin{equation}
g_{\mu\nu\,;\,\alpha}=\phi_\alpha\,g_{\mu\nu}\,,
\label{metricderivative}
\end{equation}
where a semicolon denotes a covariant derivative. 
Now, in this more general geometry, the Riemann tensor is decomposed
into two parts $R_{\alpha\beta\gamma\delta}=
K_{\alpha\beta\gamma\delta}+ T_{\alpha\beta\gamma\delta}$, where
$K_{\alpha\beta\gamma\delta}$ is the Riemann-Christoffel curvature
made of Christoffel symbols only, and $T_{\alpha\beta\gamma\delta}$ is
the $\phi$-dependent curvature.  Several other important conclusions
can be drawn from this new geometry.  Perform the following
transformations, $g_{\alpha\beta}$ goes into $\hat
g_{\alpha\beta}=f(x^\lambda)\,g_{\alpha\beta}$ and $\phi_\alpha$ into
$\hat \phi_\alpha=\phi_\alpha+\frac12 \left( \log
f\right)_{,\alpha}$, for some function $f(x^\lambda)$, 
a coma denoting simple derivative.
Then with the help of Eq.~(\ref{Gammas}) one can
work out that
$\Gamma^\alpha_{\beta\gamma}=\hat\Gamma^\alpha_{\beta\gamma}$.  This
set of transformations forms the Weyl group. Since the gauge of the
length can be changed under these transformations, but not the
transport law (given by the $\Gamma^\alpha_{\beta\gamma}$), 
one says that the geometry is invariant under gauge
transformations. Other points worth remarking is that angles between
vectors and ratios of lengths are preserved under the Weyl transport, 
and the light
cone structure too. On the other hand local lengths change as
${\hat l}^2= f(x^\lambda)l^2$.  Note also the interesting result that
if $\hat \phi_\alpha=0$ then the geometry is Riemannian. The condition
to be Riemannian is that
$\phi_{\alpha;\,\beta}-\phi_{\beta;\,\alpha}=0$ (indeed, under
closed transport the length changes by $\oint_c \frac{dl}{l}=\oint_c
\phi_\alpha dx^\alpha$, and this is zero if and only if
$\phi_{\alpha;\,\beta}-\phi_{\beta;\,\alpha}=0$).  Recall that when
$R_{\alpha\beta\gamma\delta}=0$ there is no change in direction of the
transported vector along a closed path.  Thus, the quantity
$F_{\alpha\beta}$ defined by
$F_{\alpha\beta}=\phi_{\alpha;\,\beta}-\phi_{\beta;\,\alpha}=0$ is, in
this context, analogous to the Riemann tensor
$R_{\alpha\beta\gamma\delta}$, in that when $F_{\alpha\beta}=0$ there
is no change in length of the transported vector along a closed path.
Moreover, there are further analogies between both tensors.  For
instance, the tensor $F_{\alpha\beta}$ possesses symmetries, with some
affinities to the Riemann tensor symmetries. They are,
$F_{\alpha\beta}=-F_{\beta\alpha}$, and
$F_{\{\alpha\beta;\gamma\}}=0$.

Having established a geometry in which directions and lengths have
similar behaviors in relation to transport, Weyl made the first
attempt to unify gravitation in the form of general relativity, and
electromagnetism in the form of Maxwell theory. His idea was, given
that the curvature tensor and its contractions provide a basis for a
physical picture of tidal forces and gravitation as in general
relativity, an extended geometry with its new connection and field
$\phi_\alpha$ can provide a basis for a gravitoelectromagnetic
unitary theory.  Similarly to having a Riemann geometry and
proposing a theory based on it as Einstein did for general relativity,
Weyl proposed a theory based on his own geometry. He looked for an
action, invariant under coordinate and gauge transformations, and
found that a Einstein-Hilbert term, proportional to the Ricci scalar
$R$, was not gauge-invariant and would not do. Thus, he had to resort
to an $R^2$ term, the full action being,
\begin{equation}
S=\int \left(R^2+a\,F_{\alpha\beta}F^{\alpha\beta}
\right)\sqrt{-g}\, d^4x\,,
\label{actionweyl}
\end{equation}
where $a$ is a coupling constant, and $g$ is the determinant of the
metric. Applying a careful variational procedure one finds that the
equations governing the Weyl theory are (see, e.g., \cite{adler})
\begin{equation}
G_{\alpha\beta}=8\pi\,\tau_{\alpha\beta}\,,
\label{graveqs}
\end{equation}
and 
\begin{equation}
F^{\alpha\beta}_{\quad;\beta}=j^\alpha\,,
\label{electroeqs}
\end{equation}
where $G_{\alpha\beta}$ is the Einstein tensor related to the
Christoffel connection alone, $F^{\alpha\beta}$ is the Maxwell tensor
as above, and $\tau_{\alpha\beta}$ and $j^\alpha$ are the
corresponding energy-momentum tensor and charge current, respectively,
constructed from within the theory alone. Thus, Weyl was able to
reproduce Einstein's and Maxwell's equation within a single
scheme. However, when confronted with observations the theory does not
hold, it must be rejected on fundamental physical grounds, as pointed
out first by Einstein (see, e.g., \cite{adler}). Indeed, since the
length of objects as well as intervals of time of particle
trajectories depend on the paths taken and thus on their past history,
one should observe that atoms arriving at the earth from different
distances in the cosmos would have different physical properties,
which we do not observe.  In spite of this demolishing problem, Weyl's
idea of gauging was one of the most fruitful ideas in the history of
physics. London \cite{london1927} tried first to apply the gauge ideas
of Weyl to quantum mechanics.  Then Weyl himself \cite{weyl1929}
understood that instead of gauging the metric tensor, he could gauge
the quantum mechanical wave function $\psi$ by a phase $\psi\to
\lambda\psi$ with $\lambda={\rm e}^{ie\int A_\mu dx^\mu}$ and couple
it to electromagnetism by changing the normal derivative to a
covariant derivative $\partial_\mu\to \nabla_\mu= \partial_\mu-
ieA_\mu$, where $e$ is the electric charge. These transformations
should have been called phase transformations, but due to the
similarity with Weyl's previous work the name of 1918 stuck, see
\cite{raifeartaigh} for this wonderful story.

Notwithstanding its problems in relation to unification of gravitation
and electromagnetism the door to unification schemes was open. There
is a Brazilian saying that says ``Onde passa um boi, passa uma
boiada'' (Where one ox passes a herd of oxen passes). It applies
neatly here.

\subsection{\small\bf Eddington theory (1921)}

In this conjunction, Eddington's theory was the next
\cite{eddingtonpaper} (see also \cite{eddingtonbook}).  Eddington set
forward the idea that, perhaps, the connection
$\Gamma^\lambda_{\mu\nu}$ is the primary quantity, rather than the
metric $g_{\mu\nu}$ itself.  Assuming a symmetric connection, which he
did, the Ricci tensor can be decomposed as
\begin{equation}
R_{\mu\nu}=R_{\mu\nu}^{\rm symm}+
R_{\mu\nu}^{\rm antisymm}\,,
\label{riccifromedddington}
\end{equation}
where $R_{\mu\nu}^{\rm symm}$ is the usual part of the Ricci tensor,
and
\begin{equation}
R_{\mu\nu}^{\rm antisymm} =\frac12\left(
\frac{\partial \Gamma^\lambda_{\mu\lambda}}{\partial x^\nu}
-\frac{\partial \Gamma^\lambda_{\nu\lambda}}{\partial x^\mu}
\right)\,,
\label{ricciantisymmetric}
\end{equation}
which is nonzero in general, it is zero for a metric Christoffel
connection. Then, one can identify first, the electromagnetic tensor
with $R_{\mu\nu}^{\rm antisymm}$, namely, $F_{\mu\nu}\equiv
R_{\mu\nu}^{\rm antisymm}$, and second, the new potential of the
theory with $ \Gamma^\lambda_{\mu\lambda}$, namely, $\phi_\mu\equiv
\Gamma^\lambda_{\mu\lambda}$, $\phi_\mu$ being thus the
electromagnetic potential. The metric tensor $g_{\mu\nu}$, not being
fundamental anymore, has nonetheless to be recovered. One postulates
then $g_{\mu\nu}\equiv \frac{1}{\Lambda}R_{\mu\nu}^{\rm symm}$, with
$\Lambda$ being a new fundamental constant. The line element squared,
$ds^2=g_{\mu\nu}dx^\mu dx^\nu$ is now written as,
$ds^2=\frac{1}{\Lambda}R_{\mu\nu}^{\rm symm} dx^\mu dx^\nu$. Given the
essentials of the geometry, Eddington goes on and proposes an action
of the type, $S=\int\sqrt{|R_{\mu\nu}|}d^4 x$ (this type of action was
taken up later in the Born-Infeld theory of electromagnetism
\cite{borninfeld}). It is a
theory based on an affine connection, indeed it is the first affine
theory. It is also the only one, probably due to its
awkwardness, despite its  ingeniousness. Einstein in between the years
of 1923 and 1925 fiddled with the theory, trying to find out field
equations, but could not progress (see \cite{pais}, see also
\cite{filippov} for a recent perspective on the action and equations
of Eddington's affine theory).

\subsection{\small\bf Cartan's torsion, differential geometry, 
Einstein's attempts and the
spectrum of unitary theories, and the $C$-field}

After the appearance of general relativity, differential geometry and
manifold theory started to be considered an important branch of
mathematics. Indeed, the ideas on connections and parallel transport
of Hessenberg (1917), Levi-Civita (1917), and Schouten (1917) sprang
from the establishment of the beauty and power of general
relativity (see \cite{schoutenriccikalkul} for the 
display of the new connections). These ideas were then used by Weyl (1918)
\cite{weyl1918,weylbook} and Eddington (1921) \cite{eddingtonpaper}
(see also \cite{eddingtonbook}) to propose new 
geometries and new physical theories of
gravitation and electromagnetism. In turn these theories inspired new
ways to explore theories of general connections and their properties.
For instance, Cartan in 1922 discovered the notion of torsion, which is
given by the antisymmetric part of the connection 
\cite{cartan1922,cartan1923} and from which follows 
the Riemann-Cartan geometry,
see Schouten's book \cite{schoutenriccikalkul}
(for a textbook see \cite{sabbatagasperini}).  Finally, with
the help of this paraphernalia of connections new unitary theories
were invented and proposed. 
For all these theories see the thorough book of Mme.~Tonnelat (1965)
\cite{tonnelatbook1965}, and the excellent review by Goenner (2004)
\cite{goennerreview}.

A general connection $\Gamma$ (dropping the indices, which we will do
whenever we think it is appropriate and facilitates the reading) has a
metric part as in Riemann geometry, a homothetic part as in Weyl
geometry, and a torsion as in Cartan geometry.  Thus, besides the
Riemann-Christoffel curvature, one gets a homothetic curvature, and a
torsion curvature. Unitary theories, that tried to unify the
gravitational and electromagnetic fields used one or all these new
connections and curvatures.  Let us enumerate some of these, see
\cite{tonnelatbook1965,goennerreview} for precise
citations: (i) Theories with Riemann-Christoffel and homothetic
curvatures, without torsion, i.e., $\Gamma$ is symmetric, were based
on the original one, constructed by Weyl (1918).  (ii) Theories with
Riemann-Christoffel and torsion curvatures, without homothetic
curvature, have an asymmetric $\Gamma$ which can be written as
$\Gamma=\Gamma_{\rm sym}+\Gamma_{\rm antysym}$. In their full
generality this type of theories was started by Cartan (1923), and 
the original theory, along with developments, is now 
called Einstein-Cartan theory.  In a particular case, namely, in
the case one could use the notion of distant parallelism, these
teleparallel versions were explored by Weitzenb\"ock (1925), Einstein
(1925), Infeld (1928), and others. (iii) Theories with all three
curvatures, where also $\Gamma=\Gamma_{\rm sym}+\Gamma_{\rm antysym}$,
were tried by Schouten (1924), Eyraud (1926), Infeld (1928), and
Straneo (1931).  The original theory of Eddington (1921), explored by
Einstein (1923), starts from a manifold with a connection only, the
metric being a derived entity, and follows outside this scheme,
perhaps. Einstein (1942) and Schr\"odinger (1943) even tried theories
where the fundamental tensor $g_{\alpha\beta}$ has an antisymmetric
part, $g_{\alpha\beta}=g_{\alpha\beta \;{\rm sym}}+ g_{\alpha\beta
\;{\rm antisym}}$. See \cite{tonnelatbook1965,goennerreview}.

Another idea on connections that sprang from
all these differential geometries, and is seldom
mentioned, is that the manifold can see a connection $\Gamma$ for
contravariant vectors $v$ and a different connection $\Gamma\,'$ for
covariant vectors $u$. Thus, for each connection, $\Gamma$ and
$\Gamma\,'$, one gets the usual Riemann-Christoffel curvature, a
torsion curvature, and a homothetic curvature. These two 
distinct connections
give rise to a new three-index tensor field $C$ which in turn makes
the bridge between the connections themselves, and so between the
contravariant and the covariant vectors and tensors.  The field $C$ is
defined as the covariant derivative of the identity $I$,
namely,
$C_{\alpha\beta}^{\quad\gamma}\equiv{{I_{\beta}}^{\gamma}}_{;\,\alpha}
=\Gamma_{\alpha\beta}^{\quad\gamma}
+{\Gamma\,'_{\alpha\beta}}^{\gamma}$.  For most physicists and in most
theories, this $C$ field was put to zero, probably because of its
apparent lack of physical meaning. As we will see, not for Mira
Fernandes. In the years between 1926 and 1933 he explored some of the
proposed theories by adding to them the $C$ field, while trying to
physically interpret it.

At the time there were ways, other than modifying the connection
structure of spacetime, to try unification between the gravitational
and electromagnetic fields. An important scheme
is still under study. In this scheme one sticks to
Riemann geometry and Einstein's equations (or some modifications of
these) but in spacetime dimensions higher than four $d>4$, so that the
gravitational field in the extra dimensions is seen in four dimensions
instead as an electromagnetic or some other field. Such an idea was
pursued by Kaluza (1921), Klein (1926), Einstein and Mayer (1931),
Einstein, Bargmann and Bergmann (1941), Jordan (1945) and Thiry
(1945), and Podolanski (1950), and others. These theories are
generically called Kaluza-Klein theories.

With these new ideas and connections many different schemes were
constructed \cite{tonnelatbook1965,goennerreview}.

\section{\small\bf The works of Mira Fernandes on connections and on
unitary theories of gravitation and electromagnetism}
\label{mirafernandes}

Having put forward the ideas on unitary theories of gravitation and
electromagnetism in the context of the 1920s and beginnings of 1930s
we are now ready to understand the works of Mira Fernandes, first on
connections, then on unitary theories themselves. The works on
connections \cite{mirabookoriginal,mirapaper1931} are based on the
books and papers of the mathematicians and mathematical physicists
previously referred to. The works on unitary theories are based on
ideas developed by the Italian mathematical physicist Paolo Straneo,
which in turn are based on the theories of Weyl, Eddington, Cartan,
Einstein and others already mentioned. Indeed, using Straneo's ideas
on gravitational and electromagnetic fields and their relations to
connections \cite{strlince1,strlince2,strlince3,strlince4,strnuoci},
Mira Fernandes wrote a paper in 1932 \cite{mirapaper1932}. Then Mira
Fernandes became interested in another Straneo's idea related to
teleparallel theories \cite{straneoZeitschrift} upon which he wrote a
paper in 1933 \cite{mirapaper1933}.  Let us see all these works in
detail.

\subsection{\small\bf The work on connections: (i) The book 1926 (ii)
Rendiconti  1931}

{\small\bf (i) The book 1926 ``Fundamentos da geometria diferencial dos
espa\c cos lineares'' (Foundations of differential geometry of the
linear spaces) (in Portuguese)
\cite{mirabookoriginal}.}

This book was published by the press of Museu Comercial in 1926
and has been reprinted
\cite{mirabookoriginal}. In its foreword Mira acknowledges the Dutch
mathematicians Schouten and Struik, the German mathematicians Blashke
and Weyl, and the English physicist and astrophysicist Eddington. In
his book Mira Fernandes follows Schouten's book of 1924
``Ricci-Kalk\"ul'' \cite{schoutenriccikalkul}.

After some preliminary definitions on tensors and their properties,
which take about 70 pages, the book goes on to define the linear
transport for vectors (throughout a consistent mixture of the
notation and conventions adopted by Mira Fernandes in the book and the
papers will be followed). For a contravariant vector $v^\alpha$ the
linear transport is defined as
\begin{equation}
Dv^\alpha=dv^\alpha+ \Gamma^\alpha_{\mu\beta}v^\mu\,dx^\beta\,,
\label{lineartransportcontravariant}
\end{equation}
where $D$ means covariant derivative, $d$ simple derivative,
$dx^\beta$ is the displacement vector along which $v^\alpha$ is
transported, and $\Gamma^\alpha_{\mu\beta}$ is the connection for
contravariant vectors. The linear transport for generic contravariant
tensors of any number of indices
$v^{\alpha\beta\gamma\cdot\cdot\cdot}$ can be generalized in the usual
way. In addition, the linear transport for a
covariant vector $u_\alpha$ is defined as
\begin{equation}
Du_\alpha=du_\alpha+ \Gamma\,'\,^\mu_{\alpha\beta}u_\mu\,dx^\beta\,,
\label{lineartransportcovariant}
\end{equation}
where $\Gamma\,'\,^\mu_{\alpha\beta}$ is the connection for covariant
vectors, in general different from $\Gamma\,^\mu_{\alpha\beta}$.  The
linear transport for generic covariant tensors of any number of
indices $u_{\alpha\beta\gamma\cdot\cdot\cdot}$ can also be generalized
in the usual way. A prime as a superscript will indicate from now on
quantities related to $\Gamma\,'$. The identity tensor
$I_\alpha^\beta$ has then covariant derivative given by
${I_\alpha^\beta}_{;\,\mu}={I_\alpha^\beta}_{,\,\mu} +
\Gamma^\beta_{\nu\mu} {I_\alpha^\nu} + \Gamma\,'^\nu_{\alpha\mu}
{I_\nu^\beta}$, where, of course, ${I_\alpha^\beta}_{,\,\mu}=0$,
a comma denoting simple derivative. Thus,
one can define a $C$-field, a three index tensor, as
${C_{\mu\alpha}}^\beta\equiv {I_\alpha^\beta}_{;\,\mu}$. One then has
\begin{equation}
{C_{\mu\alpha}}^\beta\equiv {I_\alpha^\beta}_{;\,\mu}=
\Gamma^\beta_{\nu\mu} {I_\alpha^\nu}
+ \Gamma\,'^\nu_{\alpha\mu} {I_\nu^\beta}\,.
\label{cfield}
\end{equation}
The tensor field $C$ links the connection for the contravariant
vectors and tensors with the connection for covariant vectors and
tensors. Due to the nonvanishing of the covariant derivative of the identity
tensor in general, one has that
\begin{equation}
\left(u_\alpha\,v^\alpha\right)_{;\,\beta}=
u_{\alpha;\,\beta}v^\alpha+u_\alpha
v^\alpha_{;\,\beta}-{C_{\beta\alpha}}^\mu u_\mu v^\alpha\,.
\label{covderivativeofcontraction}
\end{equation}
When ${C_{\beta\alpha}}^\mu=0$, the case we are used to, then the
Leibniz rule for the differentiation of a product, here a contraction,
holds, and one says, with Mira Fernandes, that the transport is
invariant by contraction.

Now, the tensor ${C_{\beta\alpha}}^\mu$ is quite general, and unwieldy
to handle, so it is of interest to simplify it, as Schouten
first suggested (see \cite{schoutenriccikalkul}). One
puts,
\begin{equation}
{C_{\beta\alpha}}^\mu=C_\beta\,I_\alpha^\mu\,,
\label{cfieldasvectorfieldincidence}
\end{equation}
i.e., the three index tensor field ${C_{\beta\alpha}}^\mu$ turns
essentially into a simple vector $C_\beta$. The derivative of a vector
contraction becomes now
\begin{equation}
\left(u_\alpha\,v^\alpha\right)_{;\,\beta}=
u_{\alpha;\,\beta}v^\alpha+u_\alpha
v^\alpha_{;\,\beta}-C_\beta\, \left(u_\alpha v^\alpha\right)\,.
\label{covderivativeofcontractionsimplifiedincidence}
\end{equation}
Mira Fernandes in the book, as well as in some of his papers, works in
$n$-dimensions, usually to be considered spacetime dimensions. He then
states that when $u_\alpha v^\alpha=0$, i.e., $v^\alpha$ belongs to
the $(n-1)$-hyperplane defined by the covector $u_\alpha$, the Leibniz
rule for the differentiation of the product is verified. When
$u_\alpha v^\alpha=0$ the vectors $u$ and $v$ are said to be incident
vectors. In this case the transport is said invariant by
incidence. Thus, there are transports invariant by contraction and
transports invariant by incidence.

Within each connection, $\Gamma$ or $\Gamma\,'$, there is an
important quantity related to the antisymmetric part of it, called
torsion. The torsion ${S_{\alpha\beta}}^\gamma$ is defined by
\begin{equation}
{S_{\alpha\beta}}^\gamma=\frac12\left(
\Gamma^\gamma_{\alpha\beta}-\Gamma^\gamma_{\beta\alpha} \right)\,,
\label{torsion}
\end{equation}
and there is an analogous definition for ${S\,'_{\alpha\beta}}^\gamma$,
\begin{equation}
{S\,'_{\alpha\beta}}^\gamma=\frac12\left(
\Gamma\,'^{\,\gamma}_{\alpha\beta}-\Gamma\,'^{\,\gamma}_{\beta\alpha} \right)\,.
\label{torsionprime}
\end{equation}
When the torsion is nonzero, the transport of a vector along a closed
path of the manifold is mapped into a transport in a
nonclosed path in the associated tangent space. There are two particular
cases of relevance. When ${S_{\alpha\beta}}^\gamma=0$ the transport is
said, in the book, contravariant symmetric. When
${S_{\alpha\beta}}^\gamma=S_{[\beta}I_{\alpha]}^\gamma$ the transport is
said contravariant hemisymmetric, a nomenclature which followed Schouten
\cite{schoutenriccikalkul}. The same applies to
${S\,'_{\alpha\beta}}^\gamma$, the torsion for the transport of
covariant vectors.

The fundamental tensor $g_{\alpha\beta}$ is the generalization of the
metric tensor, used in Riemann geometry, to more general geometries.
In general, $g_{\alpha\beta}$ can have no symmetries, the symmetric part
of it defines the lengths of vectors at a point. In the book, the
fundamental tensor $g_{\alpha\beta}$ is always considered symmetric.
Weyl geometry gives the simplest example of such a fundamental tensor.
As in Weyl geometry there is now a quantity $Q\,'_{\alpha\beta\gamma}$,
useful for
contravariant vectors $v^\alpha$, see Eq.~(\ref{metricderivative}),
defined by
\begin{equation}
Q\,'_{\alpha\beta\gamma}=g_{\beta\gamma;\,\alpha}\,,
\label{nonmetric1}
\end{equation}
where $Q\,'$ is called the nonmetricity tensor. It tells
how the fundamental tensor $g_{\alpha\beta}$ deviates
from being a pure metric tensor. Again, there are two particular
cases of relevance.
When $Q\,'_{\alpha\beta\gamma}=0$ the transport is said
contravariant metric, since in this case
$g_{\alpha\beta}$ is indeed a metric tensor 
for contravariant vectors $v^\alpha$.
When $Q\,'_{\alpha\beta\gamma}=Q\,'_\alpha
g_{\beta\gamma}$ the transport is contravariant conform, 
which is the case in Weyl theory, see Eq.~(\ref{metricderivative}).
An analogous quantity ${Q_{\alpha}}^{\beta\gamma}$ holds for covariant
vectors.
${Q_{\alpha}}^{\beta\gamma}$ is defined as
\begin{equation}
{Q_{\alpha}}^{\beta\gamma}={g^{\beta\gamma}}_{;\,\alpha}\,,
\label{nonmetric2}
\end{equation}
It tells how the raised fundamental tensor $g^{\alpha\beta}$ deviates
from being a pure metric tensor. Again, there are two particular cases
of relevance.  When ${Q_{\alpha}}^{\beta\gamma}=0$ the transport is
said covariant metric, since in this case $g^{\alpha\beta}$ is indeed
a metric tensor for covariant vectors $u_\alpha$.  When
${Q_{\alpha}}^{\beta\gamma}=Q_\alpha g^{\beta\gamma}$ the transport is
covariant conform.

With these definitions one can now express the connections $\Gamma$
and $\Gamma\,'$ in the Christoffel symbols and in the fields $C$, $g$,
$S$, $Q$, $S\,'$, and $Q\,'$ (see Eq.~(\ref{Gammas}) for the
particular case of the Weyl geometry). Indeed, it is shown in the
book, that
\begin{equation}
\Gamma^\lambda_{\alpha\gamma}=\{^{\,\,\alpha\,}_{\beta\gamma}\}
+{T_{\alpha\gamma}}^\lambda\,,
\label{Gammasagain1}
\end{equation}
\begin{equation}
\Gamma\,'^\lambda_{\alpha\gamma}=-\{^{\,\,\alpha\,}_{\beta\gamma}\}
+{T\,'_{\alpha\gamma}}^\lambda\,,
\label{Gammasagain2}
\end{equation}
where
\begin{equation}
{T_{\alpha\gamma}}^\lambda=
{C_{\gamma\alpha}}^\lambda-{T\,'_{\alpha\gamma}}^\lambda\,,
\label{GammasagainT1}
\end{equation}
and
\begin{equation}
{T\,'_{\alpha\gamma}}^\lambda=
\frac12\left( Q_{\gamma\alpha\beta}+Q_{\alpha\gamma\beta}-
Q_{\beta\alpha\gamma}\right)\,g^{\beta\gamma}
-{S\,'_{\beta\gamma}}^\nu g_{\alpha\nu}g^{\beta\lambda}
-{S\,'_{\mu\alpha}}^\nu g_{\gamma\nu}g_{\gamma\nu}g^{\beta\lambda} 
+{S\,'_{\alpha\nu}}^\lambda \,,
\label{GammasagainT2}
\end{equation}
with the last three terms involving a linear combination of the 
torsion being sometimes called the contorsion. 

Having properly defined the connections of a manifold one can go on to
define the associated curvature. The curvature of a manifold has the
same expression in terms of the connection as the Riemann-Christoffel
curvature has in terms of the Christoffel symbols connection (a
connection defined solely in terms of the metric). The expression for
the curvature for contravariant vectors is
\begin{equation}
{R_{\nu\beta\lambda}}^\alpha=
\Gamma^\alpha_{\lambda\nu;\,\beta}-
\Gamma^\alpha_{\lambda\beta;\,\nu}
+\Gamma^\alpha_{\mu\beta}\Gamma^\mu_{\lambda\nu}
-\Gamma^\alpha_{\mu\nu}\Gamma^\mu_{\lambda\beta} \,,
\label{curvature1}
\end{equation}
whereas the expression for the curvature for
covariant vectors is
\begin{equation}
{R\,'_{\nu\beta\lambda}}^\alpha=
\Gamma\,'^\alpha_{\lambda\nu;\,\beta}-
\Gamma\,'^\alpha_{\lambda\beta;\,\nu}
+\Gamma\,'^\alpha_{\mu\beta}\Gamma\,'^\mu_{\lambda\nu}
-\Gamma\,'^\alpha_{\mu\nu}\Gamma\,'^\mu_{\lambda\beta} \,.
\label{curvature2}
\end{equation}
When the curvature ${R_{\nu\beta\lambda}}^\alpha=0$ the manifold is
flat for the transport of a contravariant vector, the transport is
called contravariant parallel. When the curvature
${R\,'_{\nu\beta\lambda}}^\alpha=0$ the manifold is flat for the
transport of covariant vectors, the transport is called covariant
parallel. The curvature is also used to define a transport for a
bivector that is called contravariant equivalent in the book.  A
bivector $v^{\alpha\beta}$ is a tensor such that $v^{\beta\alpha}=-
v^{\alpha\beta}$, i.e., it is an antisymmetric two-indice tensor.
Define the quantity $V_{\nu\mu}$, as Mira Fernandes does in the book,
as $V_{\nu\mu}= {R_{\nu\mu\alpha}}^\alpha$. Then, if the transport of
$v^{\alpha\beta}$ along a closed path is zero it is called
contravariant equivalent or equiaffine.  The same rationale holds for
a covariant bivector $u_{\beta\alpha}=- u_{\alpha\beta}$.

There are particular important cases, all of them analyzed towards the
end of the book. Riemann transport is the one for which $C=0$, $S=0$,
and $Q=0$, and leads to general relativity.  Weyl transport is such
that $C=0$, $S=0$, and $Q_{\gamma\alpha\beta}= Q_\gamma
g_{\alpha\beta}$, and leads to Weyl's theory.  The so called affine
transport is such that $C=0$, $S=0$, and $Q$ is any arbitrary
quantity, like in
Eddington's theory.

Most geometries studied throughout the years have $C=0$. This is
mathematically a relief, since the field $C$ complicates the
expressions tremendously. However, for some reason, 
Mira Fernandes used manifolds in which the connections are linked
by a nonzero $C$-field, and tried to give a physical meaning to $C$,
such as an electromagnetic field in the unitary schemes he 
and others developed,
as we will see.

\vskip 0.5cm
\noindent {\small\bf (ii) Rendiconti 1931 ``Propriet\`a di alcune
connessioni lineari'' (Properties of some linear
connections) (in Italian) \cite{mirapaper1931}.}
\smallskip

\noindent
This paper \cite{mirapaper1931} shows some seven properties of
connections which are neither in Schouten's book
\cite{schoutenriccikalkul}, nor in Mira Fernandes' book
\cite{mirabookoriginal}. Let us see a typical one. Mira
assumes that the connection is invariant by incidence, i.e.,
${C_{\alpha\beta}}^\gamma=C_\alpha\,I_\beta^\gamma$, is covariant
symmetric, i.e., $S'=0$, and metric conform, i.e.,
$Q'_{\alpha\beta\gamma}=Q'_\alpha g_{\beta\gamma}$. Then, using a
result of Schouten he writes that the Riemann curvature for the
contravariant vectors ${R_{\rho\beta\lambda}}^\alpha$ and the Riemann
curvature for the covariant vectors ${R\,'_{\rho\beta\lambda}}^\alpha$
are linked through
\begin{equation}
{R_{\rho\beta\lambda}}^\alpha=
{R\,'_{\rho\beta\lambda}}^\alpha+2C_{[\rho\,;\,\beta]}
I_\lambda^\alpha \,.
\label{riemanninparticularcase}
\end{equation}
Now, contract in $\alpha$ and $\rho$ to get the link between the
Ricci tensors, 
\begin{equation}
R_{\beta\lambda}=R\,'_{\beta\lambda}+2 C_{[\lambda\,;\,\beta]}\,.
\label{ricciparticularcase}
\end{equation}
Assume further now $Q\,'_\alpha=0$, so that the connection is metric.
Then, in this case, the connection $\Gamma\,'\,^\mu_{\alpha\beta}$ is
given by the Christoffel symbols, and
${R\,'_{\rho\beta\lambda}}^\alpha={K\,'_{\rho\beta\lambda}}^\alpha$,
where ${K\,'_{\rho\beta\lambda}}^\alpha$ is the Riemann-Christoffel
curvature. Upon antisymmetrization, he finds
\begin{equation}
R_{[\beta\lambda]}=2 C_{[\lambda\,;\,\beta]}\,,
\label{ricciparticularcaseanti}
\end{equation}
since ${K\,'_{\beta\lambda}}$ is symmetric in 
$\beta\lambda$.
He now gladly proclaims, ``this formula resembles the formula 
of Eddington'', namely,
\begin{equation}
R_{[\beta\lambda]}=R\,'_{[\beta\lambda]}=
{}_{[\beta\,;}{T_{\lambda]\,\alpha}}^\alpha\,,
\label{riccieddington}
\end{equation}
where $T$ is part of the connection that is not Christoffel,
see Eqs.~(\ref{Gammasagain1})-(\ref{GammasagainT2}).
But there are differences. Eddington's theory
is an affine
theory with $C=0$, $S'=0$ and arbitrary $Q'$.

In summary, for Mira Fernandes,
${C_{\alpha\beta}}^\gamma=C_\alpha\,I_\beta^\gamma$, $S\,'=0$,
$Q\,'=0$, and one gets $R_{[\beta\lambda]}=2 C_{[\lambda\,;\,\beta]}$
and $R\,'_{[\beta\lambda]}=0$. For Eddington, $C=0$, $S\,'=0$,
arbitrary $Q\,'$, and one gets
$R_{[\beta\lambda]}={{}_{[\beta\,;}T_{\lambda]\,\alpha}}^\alpha$ and
$R'_{[\beta\lambda]}=R_{[\beta\lambda]}$. Can Mira Fernandes improve
on these similarities? Yes. Since $\Gamma+\Gamma\,'=C$, one also has
$T+T\,'=C$ (the Christoffel symbols disappear when summed). Moreover,
since here $T\,'=0$, one gets $T=C$, i.e., restoring indices,
${T_{\beta\alpha}}^\gamma=C_\alpha I_\beta^\gamma$. Contracting in
$\gamma\alpha$ he obtains, ${T_{\lambda\alpha}}^\alpha=C_\lambda$. Thus,
taking the covariant derivative yields,
$C_{[\lambda\,;\,\beta]}={}_{[\beta\,;}{T_{\lambda]\,\alpha}}^\alpha$.
Finally, using Eq.~(\ref{ricciparticularcaseanti}), he finds
\begin{equation}
R_{[\beta\lambda]}=2
{}_{[\beta\,;}{T_{\lambda]\,\alpha}}^\alpha\,,
\label{riccimira}
\end{equation}
indeed of Eddington's form (see Eq.~(\ref{riccieddington})), apart a
factor 2! This is the first instance where Mira Fernandes tries to
give a theoretical application to the $C$-field, the field that
connects the connections. The next two papers develop this idea.

\subsection{\small\bf Application to unitary theories of 
gravitation and electromagnetism: (i) Rendiconti 1932 
(ii) Rendiconti 1933}

{\small\bf (i) Rendiconti 1932 ``Sulla teoria unitaria 
dello spazio fisico"  (About the unitary theory of the 
physical space) (in Italian) \cite{mirapaper1932}.}
\smallskip

\noindent
In this paper of 1932 \cite{mirapaper1932} Mira Fernandes ventures
into unitary theories. He has already given a hint that he likes this
type of theories and speculations in the previous paper when he
mentions Eddington. But now he embraces it in full. Mira Fernandes
analyzes Paolo Straneo's papers, an Italian mathematical physicist who
belonged to the group of Levi-Civita. Straneo published papers on a
certain type of unitary theories on Rendiconti dei Lincei in the years
1931-1932 \cite{strlince1,strlince2,strlince3,strlince4} and a review
of his ideas is given in La Rivista del Nuovo Cimento in 1931
\cite{strnuoci}.

In his paper \cite{mirapaper1932}, Mira Fernandes states (I translate
freely) ``In a number of Notes published in these Proceedings 
prof.~Paolo Straneo establishes a unitary theory of gravitation and
electromagnetism, which, constituting a geometrical synthesis of the
physical phenomena, reduces itself to the theory of Einstein in the
absence of electrical phenomena''.
Straneo's connection that most interested Mira Fernandes is
\begin{equation}
\Gamma^\alpha_{\beta\gamma}=\{^{\,\,\alpha\,}_{\beta\gamma}\}
+\left(I^\alpha_\mu\psi_\nu-I^\alpha_\nu\psi_\mu
\right)\,,
\label{straneoconnection}
\end{equation}
where $\psi_\nu$ is an additional vector field of the theory, to
be equated physically to the electromagnetic potential, 
and $I^\alpha_\beta$ is the unit tensor. Equation
(\ref{straneoconnection}) is based, in a sense, on Weyl's connection,
and modifies it. However, it does not have the same mathematical
substratum, neither the same physical background. Nonetheless it was of
interest at the time. For the connection (\ref{straneoconnection}) one
can show that the curvature tensor is given by
\begin{equation}
{R_{\rho\mu\nu}}^\alpha=
{K_{\rho\mu\nu}}^\alpha+ 2 I^\alpha_\mu
\left(\psi_{\rho,\,\nu}-
\psi_{\nu,\,\rho}
\right)\,,
\label{straneocurvature}
\end{equation}
where ${K_{\rho\mu\nu}}^\alpha$ is the Riemann-Christoffel
curvature.
Contracting in $\alpha\rho$ yields
\begin{equation}
R_{\mu\nu}=
K_{\mu\nu}+ 2 
\left(\psi_{\mu,\,\nu}-
\psi_{\nu,\,\mu}
\right)\,.
\label{straneoricci}
\end{equation}
Contracting again gives 
\begin{equation}
R=K\,.
\label{straneoricciscalar}
\end{equation}
Mira Fernandes then writes Straneo's gravitational
field equation, namely,
\begin{equation}
R_{\mu\nu}-\frac12 R g_{\mu\nu} + 2\psi_{\mu\nu}=\chi 
\Theta_{\mu\nu}\,,
\label{straneofeild equations}
\end{equation}
where $\chi$ is some coupling constant, $\psi_{\mu\nu} \equiv
\psi_{\mu,\,\nu}- \psi_{\nu,\,\mu}$, and $\Theta_{\mu\nu}$ is the
energy-momentum tensor. Mira Fernandes does not write the field equation
for $\psi_\nu$, presumably it is ${{\psi_\mu}^\nu}_{;\nu}=J_\mu$, where
$J_\mu$ is some current. There is a similar theory proposed by Infeld in
1928 which is previous to Straneo's and Mira Fernandes mentions it in
passing, see also \cite{tonnelatbook1965}. Note that in a clear sense
these equations do not fulfill a scheme for full unification, as
envisaged by some 
at the time, since an energy-momentum tensor appears.

Mira Fernandes then states: ``The aim of this Note is to formulate some
considerations about the connection of Straneo, and about other
connections that lead to the same field equations and of which the
author has occupied himself in a previous paper''. This previous paper
is the Rendiconti 1931 \cite{mirapaper1931} commented above.  Then Mira
points out several things. 

To start he points out that Straneo's connection in not contravariant
metric, i.e., $Q\,'_{\alpha\mu\nu}\neq0$. Further, he takes some time to
show that assuming $Q\,'_{\alpha\mu\nu}=0$ and ${Q_{\alpha}}^{\mu\nu}=0$
there is no way one can find the curvature tensor of Straneo, a result
one could have guessed beforehand given the experience with Weyl's
connection.

He wants to go further and derive the field $\psi_\nu$ from the
connection itself! So he supposes that the connection is invariant by
incidence, covariant symmetric, and contravariant metric, in brief:
${C_{\alpha\beta}}^\gamma=C_\alpha\,I_\beta^\gamma$,
${S\,'_{\alpha\beta}}^\gamma=0$, and $Q\,'_{\alpha\beta\gamma}=0$. 
Then he finds
\begin{equation}
\Gamma^\alpha_{\beta\gamma}=\{^{\,\,\alpha\,}_{\beta\gamma}\}
+{T_{\beta\gamma}}^\alpha
\,,
\label{miraconnection1}
\end{equation}
\begin{equation}
\Gamma\,'^\alpha_{\beta\gamma}=-\{^{\,\,\alpha\,}_{\beta\gamma}\}
+{T\,'_{\beta\gamma}}^\alpha
\,,
\label{miraconnection2}
\end{equation}
with 
\begin{equation}
{T_{\beta\gamma}}^\alpha=C_\gamma\, I_\beta^\alpha
\,,
\label{Tmira1}
\end{equation}
\begin{equation}
{T\,'_{\beta\gamma}}^\alpha=0
\,.
\label{Tmira2}
\end{equation}
Now, in his previous Rendiconti \cite{mirapaper1931} he displayed
\begin{equation}
{R_{\alpha\beta\gamma}}^\delta=
{R\,'_{\alpha\beta\gamma}}^\delta+ 2 
C_{[\alpha\,;\,\beta]}I_\gamma^\delta.
\label{curvaturecurvaturedah}
\end{equation}
For the connection under study one has,
${R\,'_{\alpha\beta\gamma}}^\delta= {K\,'_{\alpha\beta\gamma}}^\delta$,
the Riemann-Christoffel tensor. One can show without difficulty that 
${K\,'_{\alpha\beta\gamma}}^\delta={K_{\alpha\beta\gamma}}^\delta$,
i.e., Riemann-Christoffel tensor for covariant vectors is the same as 
the Riemann-Christoffel tensor for contravariant vectors.
Contracting in $\alpha\delta$, 
and noting that since $S'=0$ one has
$C_{[\alpha\,;\,\beta]}=C_{[\alpha\,,\,\beta]}$, he finds
\begin{equation}
R_{\beta\gamma}=
K_{\beta\gamma}+  \left(C_{\gamma\,,\,\beta}-
C_{\beta\,,\,\gamma}\right)\,.
\label{miraricci}
\end{equation}
Contracting again gives 
\begin{equation}
R=K\,.
\label{miraricciscalar}
\end{equation}
Comparing Straneo's equation, 
Eq.~(\ref{straneoricci}), with Mira Fernandes' equation, 
Eq.~(\ref{miraricci}), it is clear that Straneo's Ricci tensor is recovered
if one puts
\begin{equation}
C_\mu=-2\psi_\mu\,.
\label{recoverofstraneobymiramiraricciscalar}
\end{equation}
Thus, the field $C$, that links the distinct connections for
contravariant and covariant vector fields, provides the 
electromagnetic field potential $\psi$. It is perhaps the first
instance that the field $C$ receives a physical interpretation. 

Moreover, the field $C$, and thus the electromagnetic field $\psi$, is
also related to both the torsion and the nonmetricity tensors.
The link with the torsion goes as follows: 
since quite
generally ${C_{[\alpha\beta]}}^\gamma=
{S_{\beta\alpha}}^\gamma +{S\,'_{\beta\alpha}}^\gamma$, and here
$S\,'=0$,
one finds ${S_{\beta\alpha}}^\gamma=S_{[\alpha}I_{\beta]}^\gamma$,
with $S_\alpha\equiv C_\alpha$, which means, in the nomenclature of
Schouten \cite{schoutenriccikalkul} and Mira Fernandes 
\cite{mirabookoriginal}, that the connection is contravariant
hemisymmetric.
The link with the nonmetricity tensor can be easily seen
since for ${C_{[\alpha\beta]}}^\gamma= C_\alpha I_{\beta}^\gamma$
one finds that $Q\,'_{\alpha\beta\gamma}=-g_{\beta\mu}
g_{\gamma\nu}{Q_\alpha}^{\mu\nu}+ 2C_\alpha g_{\beta\gamma} $. Since
here $Q\,'=0$ one has $Q_{\alpha\beta\gamma}=Q_\alpha g_{\beta\gamma}$
with $Q_\alpha=2C_\alpha$, providing the link. The connection
is contravariant
conform, of Weyl type.  A final property referred in the paper
comes from the fact that 
a contraction
in $\gamma\delta$ implies that $V\,'_{\alpha\beta}=0$,  so the
connection is covariant equivalent or equiaffine.

For a four-dimensional spacetime, $n=4$, the connection, in Mira
Fernandes' words, satisfies ``all the
conditions attributed by Straneo to the structure of the physical
space''. The vector $\psi_\mu$ representing the electromagnetic
potential is such that
\begin{equation}
\psi_\mu=-\,\frac12 \,C_\mu
\,,
\label{psiasc}
\end{equation}
and the torsion and nonmetricity quantities are given by
\begin{equation}
S_\mu=\frac12\,Q_\mu=C_\mu\,.
\label{psiasc2}
\end{equation}
The field $C$ provides much of the physical and geometrical content of
the analysis. 

Summarizing, the linear connection proposed by Mira Fernandes is
metric contravariant but not metric covariant. It is contravariant and
covariant conform, i.e., conserves the angles and the ratios of
lengths in both transports and preserves lengths only in the
contravariant transport. It also conserves covariant bivectors and
covariant p-vectors (totally antisymmetric tensors with any number of
indices) in the transport. Mira Fernandes concludes the section stating
that the connection ``is distinct from Weyl since
${C_{\alpha\beta}}^\gamma\neq0$''. The connection is invariant by
incidence, not by contraction. When $C_\alpha=C_{,\,\alpha}$, i.e,
$C_\alpha$ is the gradient of some function $C$, then one recovers
Weyl. In the rest of the paper he does a similar analysis for a
connection that is invariant by incidence, contravariant symmetric,
and covariant metric, i.e.,
${C_{\alpha\beta}}^\gamma=C_\alpha\,I_\beta^\gamma$,
${S_{\alpha\beta}}^\gamma=0$, and ${Q_{\alpha}}^{\beta\gamma}=0$. It
is of no great use to repeat this analysis here since it is much the
same as the previous one.

\vskip 0.5cm
\noindent
{\small\bf (ii) Rendiconti 1933 ``Sulla teoria unitaria 
dello spazio fisico"  (About the unitary theory of the 
physical space) (in Italian) \cite{mirapaper1933}.}
\smallskip

\noindent
This paper of Mira Fernandes \cite{mirapaper1933} has the same title
as the previous one \cite{mirapaper1932}, which was perhaps a common
practice in the Rendiconti when the author wrote on the same 
subject.
In this paper, as usual, he states the definitions of 
the fields $C$, $S$, $S'$,
$Q$, $Q\,'$, $\Gamma(g,T)$, $\Gamma\,'(g,T\,')$, $T(Q,S,g)$,
$T\,'(Q\,',S\,',g)$,
where again we have dropped the indices.  Some properties related 
to these quantities are
${T_{[\alpha\gamma]}}^\lambda ={S_{\alpha\gamma}}^\lambda$,
${C_{[\alpha\gamma]}}^\lambda ={S_{\gamma\alpha}}^\lambda+
{S\,'_{\gamma\alpha}}^\lambda$, and $Q\,'=Q'(S,Q,g)$ a complicated
function which we do not need to show here explicitly.
He takes from his book \cite{mirabookoriginal} the relation
between the Riemann $R\,'$ and the Riemann  $R$,
\begin{equation}
{R\,'_{\alpha\beta\gamma}}^\delta=
{R_{\alpha\beta\gamma}}^\delta
+ 2\,{S\,'_{\alpha\beta}}^\mu {C_{\mu\gamma}}^\delta
+ {{C_{\beta\gamma}}^\delta}_{,\,\alpha}
-{{C_{\alpha\gamma}}^\delta}_{,\,\beta}
\,,
\label{relationR'R}
\end{equation}
and the relation between the Riemann $R$ and the Riemann-Christoffel 
$K$,
\begin{equation}
{R_{\alpha\beta\gamma}}^\delta=
{K_{\alpha\beta\gamma}}^\delta
+ {{T_{\gamma\alpha}}^\delta}_{;\,\beta}
-{{T_{\gamma\beta}}^\delta}_{;\,\alpha}
+ {T_{\gamma\beta}}^\mu\,{T_{\mu\alpha}}^\delta
- {T_{\gamma\alpha}}^\mu\,{T_{\mu\beta}}^\delta\,.
\label{relationRK}
\end{equation}
A similar relation between ${R\,'_{\alpha\beta\gamma}}^\delta$ and 
the other prime quantities also holds.
Suppose now that the connection is metric covariant
${Q_\alpha}^{\beta\gamma}=0$, and that $T$ is antisymmetric in the
lower indices $\alpha\gamma$ so that ${T_{\alpha\gamma}}^\lambda=
{S_{\alpha\gamma}}^\lambda$. Then ${T_{\alpha\gamma}}^\gamma=0$, with no
sum in the repeated indices.
Suppose further that spacetime is flat for contravariant vectors, 
i.e., ${R_{\alpha\beta\gamma}}^\delta=0$.
Then, from the symmetries of ${R_{\alpha\beta\gamma}}^\delta$
and the properties mentioned above, one finds
upon using (\ref{relationRK}) that
\begin{equation}
{K_{\alpha\beta\gamma}}^\delta
-{{T_{\alpha\beta}}^\delta}_{;\,\gamma}
+{T_{\alpha\beta}}^\mu\,{T_{\mu\gamma}}^\delta
=0\,.
\label{fundamentalequation}
\end{equation}
Mira Fernandes states that these equations ``are the fundamental
equations of the unitary theory of prof.~Straneo''. Unfortunately, he
does not quote the paper in which Straneo writes these equations. They
are not in Rendiconti \cite{strlince1,strlince2,strlince3,strlince4}
neither in Nuovo Cimento \cite{strnuoci}. It is most certain that Mira
Fernandes is referring to a paper of Straneo (in German) ``Einheitlich
Feldtheorie der Gravitation und Elektrizit\"at'' published in
Zeitschrift f\"ur Physik in 1932 \cite{straneoZeitschrift}. The
equations (\ref{fundamentalequation}) define an absolute transport for
covariant vectors. Straneo recovers distant parallelism of Cartan,
Weitzenb\"ock, Einstein, and others
\cite{tonnelatbook1965,goennerreview}. With a somewhat pompous stance,
Mira Fernandes then states (my translation): ``The above equation
translates a remarkable structure of physical space characterizing a
chronotope of contravariant curvature zero and metric covariant.''
Also, since here ${\Gamma_{\mu\nu}}^\alpha=
-{\Gamma\,'_{\mu\nu}}^\alpha$, one has that ${C_{\mu\nu}}^\alpha=0$
and $Q\,'_{\alpha\beta\gamma}=0$, so that
${R\,'_{\alpha\beta\gamma}}^\delta=0$. There is also distant, or
absolute, parallelism for covariant transport.

Mira Fernandes then turns again to the tensor
${C_{\alpha\beta}}^{\gamma}$ and shows that if this is nonzero then
the equations of Straneo still hold for contravariant vectors, but now
${R\,'_{\alpha\beta\gamma}}^\delta$ can be nonzero, i.e., there is
absolute transport for contravariant vectors but not for covariant
vectors. Mira's final remark is: ``E non sar\`a privo d'interesse, per
future utillizzazioni della teoria unitaria l'aver constatato che le
equazzioni del prof.~Straneo sono compatibili con connessioni lineari
(in numero infinito) in cui il tensore $({C_{\alpha\beta}}^\gamma)$
non \`e nullo; ci\`o che non sono invariante per contrazioni'', i.e., 
he states (my
translation): ``And it will be not without interest, to a future use of
the unitary theory to have ascertained that the equations of
prof.~Straneo are compatible with linear connections (in infinite
number) in which the tensor $({C_{\alpha\beta}}^\gamma)$ is nonzero;
i.e., that they are not invariant by contraction''. In this way Mira
Fernandes tries to push forward once again the field $C$ that links
the connections. However, this time, there is no attempt, not even
indirectly, to relate it to the electromagnetic potential.

\section{Conclusions}
\label{conclusions}

\subsection{\small\bf What else?}

There are other related publications which I did not delved into and
certainly deserve a closer scrutiny.

In 1924 Mira Fernandes published his first book, ``Elements of the
theory of the quadratic forms'' (in Portuguese) \cite{book1924}. It
is divided into two parts: Algebraic forms and Differential forms. It
is self-contained, written at a somewhat advanced level but of easy
reading. It would be of interest to know which are the sources Mira
Fernandes used to write this book.

In 1934 there is another publication in Rendiconti on unitary theories
with the title ``The unitary theory of physical space and the
relativistic equations of atomic mechanics'' (in Italian)
\cite{paper1934}. It is a paper on Dirac's equation and tries to unify
general relativity and wave mechanics. As remarked in
\cite{costaleitegagean} it is a work that certainly deserves
interpretation in a historical context.

In 1945 and in 1950 Mira Fernandes published two papers in Portugaliae
Mathematica in order to develop fresh Einstein work on bivectors
\cite{eins1,eins2}, where Einstein tried to find fundamental equations
without the use of differential equations.  The first paper of the set
has the the title ``Finite connections'' (in Italian)
\cite{paper1945}, and the second ``Finite transports'' (in Italian)
\cite{paper1950transp}.

In 1950, in Revista da Faculdade de Ci\^encias, Mira Fernandes
published a paper with the title ``The geodesics of the unitary
space'' (in Italian) \cite{paper1950geodcompl}. The paper is on
complex manifolds, generalizing results of Coburn, an American
mathematician, and it has nothing to do with the papers on the unitary
theories of the physical space. This paper of Mira Fernandes is quoted
in the book ``Ricci Calculus'' \cite{schouten1954}, the 1954 
second edition, now in English, of Schouten,
a citation that must have given him much
pleasure. The paper is also quoted in the book of 1955 of
Mme.~Tonnelat on the unitary theories of Einstein and Schr\"odinger
\cite{tonnelat1955}.

There are other papers of Mira Fernandes on differential geometry of
much interest but they are outside the main theme of our work on the
extension of general relativity into unitary theories of gravitation
and electromagnetism using the full possibilities offered by the
connections of differential geometry. One that perhaps is worth
quoting, published in 1932, is about the brachitochronous problem of
Zermelo \cite{mirazermelo}, which, in turn, can be put in a Finsler
geometric context, as has been shown recently, see \cite{herdeiro}.

\subsection{\small\bf The fate of unitary theories}

In the 1920s and beginnings of the 1930s the only two fields known
were the gravitational and electromagnetic fields, assumed to be
classical in the proposed unitary theories. Since then two more
fields have been discovered, the weak and strong fields, and these,
together with the electromagnetic field, have proved to be quantum
fields. The mere existence of these two additional fields already puts
in jeopardy the program of unifying the gravitational and
electromagnetic fields alone. The fact that the fields are quantum in
character dismisses definitely the whole program, based on a classical
setup. Nonetheless, there are important ramifications taken out from
the unification idea.

First, although the theories which change Riemann geometry, as those
used by Mira Fernandes, are not in fashion nowadays as theories of
unification of gravitation and electromagnetism, some of them were
reverted to theories that embody gravitation, torsion, energy-momentum
and elementary spin, and are called Einstein-Cartan theories, or
Einstein-Cartan-Kibble-Sciama theories, the latter two names appearing
because they showed first that the Einstein-Cartan theory can be
formulated as a gauge theory with local Poincar\'e invariance in flat
spacetime, see, e.g., \cite{hehl} (see also \cite{sabbatagasperini}).

Second, the idea of unification still persists but on a different
basis. The electromagnetic field, and its associated massless quantum
particle, the photon, has been already joined with the weak field to
produce the electroweak field. There remains the possibility that all
three fields, electromagnetic, weak and strong, can be unified in a
grand unified theory with their associated massless quantum particles.
One can then hope that the gravitational field with its associated
massless graviton, and the grand unified field and particles, can be
united in an ultimate theory of unification. The most current
celebrated theories make use of the gravitational field in extra
dimensions in order to try to obtain, in four dimensions, the
gravitational field itself and the grand unified field (which itself
contains and generalizes 
the electromagnetic field sought for in the early attempts).
These theories are reminiscent of the ideas of Kaluza and Klein back
in the 1920s, that were then used in their na\"ive form by Einstein
and others \cite{tonnelatbook1965,goennerreview}, but not touched or
mentioned by Mira Fernandes.  Nowadays these theories are generically
called Kaluza-Klein theories.  They were incorporated into
supergravity \cite{modernkk}, and then reappeared in string theory in
a prominent form, see, e.g., \cite{polch}.

The name of such theories has been changing, unitary theories at
first, then unified field theories, and nowadays theories of
everything.  Will their fate be the same as Mie's theory?

\vskip 0.1cm
\noindent
-----------------

\vskip 0.5cm
{\small

\subsection*{\small\bf Sources}

I have benefited from several sources which helped me to put the works
of Mira Fernandes in context. 

\vskip 0.1cm
\noindent  Direct sources:

$\cdot$Schouten 1924 \cite{schoutenriccikalkul}. Jan Arnoldus Schouten
was Dutch and, at the time, to write in German gave a much wider
audience. The book Ricci-Kalk\"ul \cite{schoutenriccikalkul} is
written in German. I have used this book to connect the formulas in
Mira Fernandes' book \cite{mirabookoriginal} with the
formulas in Schouten's book \cite{schoutenriccikalkul}.

$\cdot$Mira Fernandes' works 1926-1933
\cite{mirabookoriginal,mirapaper1931,mirapaper1932,
mirapaper1933}. Mira Fernandes works on differential geometry along
with other works have been now reprinted by Gulbenkian Foundation in
three volumes \cite{gulb1,gulb2,gulb3}.  Prior to the publication of
volumes 2 and 3, several papers, including the ones published in
Rendiconti in the year 1931 to 1933 commented above
\cite{mirapaper1931,mirapaper1932,mirapaper1933}, were facilitated to
me by the staff in {\tiny CEMAPRE} (Centre for Applied Mathematics and
Economics) of {\tiny ISEG} (Instituto Superior de Economia e
Gest\~ao), where copies of all the works and papers of Mira Fernandes
are kept. ISEG is one of the places where Mira Fernandes taught.

$\cdot$Straneo's papers 1931-1932
\cite{strlince1,strlince2,strlince3,strlince4,strnuoci,straneoZeitschrift}.
These papers were essential for our analysis, since Mira Fernandes
bases his works on unitary theories on them. Paolo Straneo was a
mathematical physicist from Genoa. Levi-Civita presented to the
Academy of Lincei papers on unitary theories from Paolo Straneo,
Attilio Palatini, Pia Nalli, Mira Fernandes and others. Initially, I
have had access to his review paper published in 1931 in La Rivista
del Nuovo Cimento \cite {strnuoci} and to a paper published in 1932 in
Zeitschrift f\"ur Physik \cite{straneoZeitschrift} only. Now, many of
his papers are on the internet.

\vskip 0.1cm
\noindent  Indirect sources:

$\cdot$Gagean and Leite 1990 \cite{costaleitegagean}. Gagean and Leite
wrote a remarkable article 
\cite{costaleitegagean}. In this article there are sections devoted to
Mira Fernandes where a historical description in context of Mira
Fernandes' work and stance is given. It was the first article
on Mira Fernandes I came across.

$\cdot$Adler, Bazin, Schiffer 1965 \cite{adler}. This book is a master
piece. It is perhaps the first text book in general relativity written
from a physical point of view, and superbly so. The other previous
good text books were written by mathematicians, even Eddington's book
\cite{eddingtonbook},
its title says it all.  Adler's book contains a chapter in which Weyl's
theory is masterly explained, and it should not be missed by the
interested reader. In the second edition of
1974 the chapter is maintained.

$\cdot$Pais 1982 \cite{pais}. This is a tour de force biography, all
scientific work of Einstein is reviewed. Although there are some flaws,
and understandably so given the huge scope and commitment of the book,
the part concerned with unified field theories is fantastically clear.

$\cdot$Tonnelat 1965 \cite{tonnelatbook1965}. This book
\cite{tonnelatbook1965} is very important in our context. It makes a
thorough review up to 1965 of the whole sets and ramifications of
unified theories in vogue and out of fashion. It does not quote
Mira Fernandes works on unitary theories.  However, it quotes
works by Straneo and Infeld (Straneo's in a footnote), helping tremendously
to connect Mira Fernandes to the main stream of the time. Without
Mme.~Tonnelat's book it would have been much more difficult
to put Mira Fernandes' work on unitary theories in context.

$\cdot$Goenner 2004 \cite{goennerreview}. This is a review paper on
unified field theories, the first part of it up to the beginning of the
1930s \cite{goennerreview} and the second part yet to be published. This
work connects smoothly to Tonnelat's book of 1965
\cite{tonnelatbook1965}, although it is clear that the author has done
thoroughly independent work. It also quotes Straneo's work, and so also
helps in connecting Mira Fernandes' works to the main stream.

\medskip
For a shorter version of this article see \cite{lemos}.

%\vskip 2cm
%\noindent 
%{\bf Acknowledgments} --
\section*{\small\bf Acknowledgments}

It is hard to write on others' work, specially from a historical
standpoint, because first one has to understand the overall context in
which the works have been written, and second one has to understand
the works themselves as well as their motivations.  This task of
trying to understand Mira Fernandes' works on unitary theories,
differential geometry and general relativity, took me more than a year.
Of course, on one hand I had to be motivated to do it, on the other
hand several people have helped.

My motivations to understand Mira Fernandes' work started while I was
in in Cambridge University during my Ph.D. times. In 1983, in Galloway
and Porter bookshop in Sidney Street, a second hand bookshop, I came
across the book ``Relativity: the general theory'' by Synge
\cite{syngegeneral}. I turned the pages and at the end in the
bibliography I hit upon the name of Mira Fernandes, his paper of 1932
published in Rendiconti \cite{mirapaper1932} was cited.  I was
surprised and glad, seeing a Portuguese scientist being quoted,
posthumously, in such an important book. The book was expensive but I
bought it immediately with no more qualms or regrets. A further
motivation came from Manuel Sande Lemos, my grandfather, who lived 102
years and managed to cross three centuries. He got the degree of
chemical engineer at Instituto Superior T\'ecnico {\tiny (IST)} in the
beginning of the 1920s, when IST was at Rua do Instituto Industrial,
Conde Bar\~ao. He attended Mira Fernandes courses on Differential and
Integral Calculus and on Rational Mechanics, having a great admiration
for him. He was always telling me about him, and advising me that I
should study his works, specially the equations named Mira
Fernandes. I still don't know which equations are these, perhaps he
also did not know. When the opportunity came I had no doubts I should
embrace this project of studying his works on unitary theories.

I was helped by several people to whom I grateful thank. Nuno Crato
has pushed me into giving the talk, and Jo\~ao Teixeira Pinto and Luis
Saraiva have invited me to give it in the ``Mira Fernandes e a sua \'epoca,
Historical Conference in honor of Aureliano de Mira Fernandes
(1884-1958)'' realized in 2009 at IST, this article
being in the Proceedings of it. Vera Lameiras, secretary of 
Cemapre-ISEG, has helped to find Mira's papers in their archives.
Although I do not know them personally, those who have collected Mira
Fernandes papers and made his complete works possible, namely, Jaime
Campos Ferreira, Luis Canto Loura, Joaquim Moura Ramos, Dulce Cabrita,
Vicente Gon\c calves, and others certainly, have made a 
significant contribution to put in real perspective Mira Fernandes'
work. Manuel Fiolhais has sent me from the Library 
of the University of Coimbra two papers of
Straneo \cite{strnuoci,straneoZeitschrift} which helped 
in the
analysis of the whole context.
}

\vskip 0.1cm
{\small This work was partially supported by FCT - Portugal 
through projects \hfill\newline
CERN/FP/109276/2009 and PTDC/FIS/098962/2008.}

\end{document}